\documentstyle[aps,twocolumn,floats,prl]{revtex}
\begin{document}
\thispagestyle{empty}
\draft
\title{On the rotation of the universe}
\author{Yuri N. Obukhov$^{1,2}$, Vladimir A. Korotky$^{2,3}$, and 
Friedrich W. Hehl$^1$}
\address{$^1$Institute for Theoretical Physics,
University of Cologne, D-50923 K{\"o}ln, Germany}
\address{$^2$Department of Theoretical Physics, Moscow State University,
117234 Moscow, Russia}
\address{$^3$Department of Mathematics, Yaroslavl State Technical University,
150023 Yaroslavl, Russia}
%\address{\sl Submitted to Physical Review Letters}
\maketitle
%\pacs{PACS numbers: 98.80.Es, 41}

Recently Nodland and Ralston \cite{nod} have analyzed the data on 
polarized electromagnetic radiation from distant radio sources. The dipole 
effect of the rotation of the plane of polarization was reported in
the form $\beta={1\over 2}\Lambda^{-1}_s\,r\,\cos\gamma$, where $\beta$
is the residual rotation angle between the polarization vector and the 
direction of the major axis of a source (remaining after the Faraday 
rotation is extracted), $r$ is the distance to the source, $\gamma$ is the
angle between the direction of the wave propagation and the constant 
vector $\vec{s}$. The analysis of data for 160 radio sources yielded 
\cite{nod} the best fit for the constant $\Lambda_s=(1.1\pm0.08)\,{2 \over 3}
\,{10^{15}\over H}\,$m yr$^{-1}$ (with the Hubble constant $H$), and the 
$\vec{s}$-direction RA\,(21h$\pm$2h), dec\,($0^\circ\pm20^\circ$).

In an attempt to explain their observations, Nodland and Ralston  
concluded that it is impossible to understand such an effect within 
conventional physics. Instead, they considered a modified electrodynamics
with the Chern-Simons-type term violating Lorentz invariance \cite{car},
and related $\Lambda_s$ to the coupling constant of that term.

In this comment we want to point out a different explanation: global 
cosmic rotation. It is of purely geometrical origin and is completely within 
conventional physics. Quite early \cite{rot1}, cosmological 
models with rotation (and, in general, with shear) attracted considerable 
attention. For the {\it mixed} effects of global vorticity and shear strong 
limits are known (see, e.g., \cite{haw}) on the value of rotation. However,
shear and rotation manifest themselves differently in observations. If one 
carefully distinguishes them, one finds substantially weaker limits on the 
cosmic vorticity \cite{grg,jetp}. 

Rotation of polarization of an electromagnetic wave is a typical effect of 
the cosmic rotation. Actually, not only polarization but, in general, all 
other optical characteristics of an image (size, shape, orientation) are 
affected by the curvature of spacetime \cite{sachs}. The specific effect 
reported by Nodland and Ralston, and earlier by Birch \cite{birch}, can be
extracted from the geometry of bundles of null geodesics (rays) as follows: 
After solving the null geodesics equations, one can construct a Newman-Penrose 
null frame $\{l,n,m,\overline{m}\}$ is such a way that $l$ coincides with the 
wave vector $k$, and the rest of the vectors are covariantly constant along 
$l$. Then we can identify a polarization vector with, e.g., $m$. If we now 
consider the deformation of a source's image {\it with respect to that frame},
the angle between the major axis of an image and $m$ will be exactly the
observable relative angle $\beta$. 

For a wide class of viable spatially homogeneous rotating models 
\cite{grg,jetp}, combining the knowledge of explicit solutions for the null 
geodesics equations with the Kristian-Sachs expansion technique, one finds
\vspace{-0.2cm}\begin{equation}
\beta=\omega\,r\,\cos\gamma + O(Z^2),
\end{equation}
where $\omega$ is the present magnitude of the rotation and $\gamma$ is the 
angle between the direction to a source and the direction of the cosmic 
vorticity. Higher angular corrections quadratic in the red shift $Z$ are not 
displayed. Recalling \cite{nod} the value of $\Lambda_s$, we immediately find 
\vspace{-0.1cm}\begin{equation}
{\omega\over H}=6.5\pm 0.5.\label{omega}
\end{equation}
This is larger than the estimate \cite{grg} obtained earlier on the basis of 
Birch's data \cite{birch}. Also the direction of $\vec{s}$, which we now 
interpret as the direction of the cosmic vorticity, is different from the 
older estimate RA\,(12h$\pm$2h), dec\,($-35^\circ\pm30^\circ$), based on the
observations of Birch \cite{grg,jetp}. However, within the error 
limits, the two directions are orthogonal to each other. 

We hope that the new data of Nodland and Ralston may provide, as compared to 
Birch's results, a substantially improved estimate of the magnitude and 
direction of the cosmic rotation. It is worthwhile to stress again that our
explanation of the dipole anisotropy in observations of polarization of radio
sources is {\it within} conventional general relativity. The value 
(\ref{omega}) is {\it not} in conflict with other observational data (in 
particular, not with the limits of anisotropy of the microwave background 
radiation), see \cite{grg,jetp}. Cosmological rotation may be significant
for models of galaxy formation \cite{silk}.

Here we do not intend to enter the discussion \cite{stat} on the statistical 
significance of the results of Nodland and Ralston. Evidently, further careful 
observations and statistical analyzes will be extremely important in
establishing the true value or finding upper limits for cosmic rotation.

\end{document}